\documentstyle[12pt]{article}
\begin{document}
\newcommand{\be}{\begin{equation}}
\newcommand{\ee}{\end{equation}}
\newcommand{\barr}{\begin{eqnarray}}
\newcommand{\earr}{\end{eqnarray}}

\begin{titlepage}
\title{The Role of $\Delta$(1232) in Two-pion Exchange Three-nucleon
       Potential}
\author{ K. Kabir, T. K. Dutta\thanks{Current address: Atomic Energy
         Commission, Savar, Dhaka}, Muslema Pervin\thanks{Current
         address: Dept. of Physics, Southern Illinois University,
         Carbondale, IL62901, USA}, and
         L. M. Nath\thanks{Email address: physics@du.bangla.net} \\
         Department of Physics, University of Dhaka  }
\maketitle
\vspace{1.2cm}
\begin{abstract}
In this paper we have studied the two-pion exchange three-nucleon potential
$(2\pi E-3NP)$ using an approximate $SU(2) \times SU(2)$ chiral symmetry
of the strong interaction. The off-shell pion-nucleon scattering amplitudes
obtained from the Weinberg Lagangian are supplemented with the
contributions from the well-known $\sigma$-term and the $\Delta(1232)$
exchange. It is the role of the $\Delta$-resonance in $2\pi E-3NP$,
which we have investigated in detail in the framework of the Lagrangian
field theory. The $\Delta$-contribution is quite
appreciable and, more significantly, it is dependent on a parameter $Z$
which is arbitrary but has the empirical bounds $|Z| \leq 1/2$. We find that
the $\Delta$-contribution to the important parameters of the
$2\pi E-3NP$ depends on the choice of a value for $Z$, although the
correction to the binding energy of triton is not expected to be very
sensitive to the variation of $Z$ within its bounds.
\end{abstract}
\setcounter{page}{1}

\end{titlepage}

\section{Introduction}

It is well-known that two-nucleon potentials are not always adequate in
explaining nuclear properties. For example, all realistic two-body
potentials which fit the two-nucleon data quite well, fail to reproduce the
binding energy of triton~\cite{Stadler,Friar}. 
The experimental binding energy of $^3H$ is 8.48 MeV,
while calculations with the well-known two-body local potentials
fall short by 0.5 -- 1.25 MeV. An obvious attempt to overcome the deficiency
is to include the three-nucleon potential (3NP) in binding energy
calculations. Computational techniques for trinuclear systems with the
inclusion of three-nucleon potentials have become sufficiently mature to make
such attempts worthwhile~\cite{Stadler,Ishikawa}. 
Because of the short-range two-body repulsion between the nucleons
tending to keep them apart, we expect that the two-pion exchange
three-nucleon potential ($2\pi E-3NP$) will have a larger effect than
the relatively shorter range contributions to the 3NP due to the exchange
of heavier mesons.

To construct the $2\pi E-3NP$ we need the pion-nucleon scattering amplitudes
with the pions off-mass-shell. An important mechanism in $\pi N$ scattering
is the formation of the $\Delta $-resonance. We study the effect of the 
$\Delta (1232)$ by considering the most general form of the $\pi N\Delta $
interaction Lagrangian~\cite{Nath} which
has been applied extensively in low energy $\pi N$ scattering \makebox{[5 - 8]}
and photo- and electro-production of
pions~\cite{photo}.
This Lagrangian contains a parameter $Z$ whose value is arbitrary. However,
low energy phenomenology [5-9] constrains $Z$ to lie between -1/2 and 1/2.
The other pieces of the effective $\pi N$ interaction Lagrangian have been
obtained from the nonlinear chiral Lagrangian of Weinberg~\cite{Weinberg1}. 
The Weinberg Lagrangian incorporates the nucleon-exchange effects in $\pi N$
scattering and, in addition, there is either a direct $\pi \pi NN$
interaction, or $\pi N$ scattering via $\rho $-exchange. Furthermore, we
have added in the pion-nucleon $\sigma $-amplitude, $A_\sigma ^{(+)}$,
parametrized in an appropriate manner, to account for some well-known
constraints \makebox{[11 - 13]} in the scattering amplitude $A^{(+)}$,
which follow from Current Algebra and
Partial Conservation of Axial-vector Current. The parameters of
$A_\sigma ^{(+)}$ have been adjusted by using the recent
information on the amplitude $\bar{F}^{(+)}$ in the subthreshold region,
obtained by analyzing the data from meson factories~\cite{sub}.
The model for pion-nucleon interaction so constructed is also compatible
with low-energy $\pi N$ data.

The two-pion exchange three-nucleon potential constructed from our model of 
$\pi N$ interaction is dominated by the $\Delta $-resonance and hence depends
on $Z$. The nonlinear realization of chiral symmetry proposed by
Weinberg~\cite{Weinberg1} 
leads to a pseudovector $\pi NN$ coupling which does not contribute to the 
$2\pi E-3NP$ in the appropriate non-relativistic limits. The contribution to
the $2\pi E-3NP$ from the direct $\pi \pi NN$ interaction or from the
$\rho $-exchange is small compared to the contribution from
$\Delta $-exchange.

Our purpose in this work is to examine in detail whether the parameter
$Z$ in the $\pi N\Delta$ interaction Lagrangian introduces appreciable
Z-dependence in the three-nucleon potential and, consequently,
in the calculations of physical quantities like the binding energy of triton. The three-nucleon potential
obtained from our model is of the same form as the Tucson-Melbourne (TM)
potential~\cite{TM} or the Brazil potential~\cite{Brazil}, 
which contains four parameters a, b, c and d. In the present case
b and d are functions of $Z$. The parameter b, which gives
the dominant contribution to the binding energy correction of triton,
is not very sensitive to $Z$, although the $\Delta$-contribution
to b, $b_{\Delta}$, varies appreciably with $Z$.
The reason for the insensitivity of b to variations of $Z$ is that,
in our model, the amplitude $A_{\sigma}^{(+)}$ is also indirectly Z-dependent
through the slope parameter $\sigma^{\prime}$ \mbox{(Sec. \ref{secsigma})}.
The $\Delta$-exchange and the amplitude $A_{\sigma}^{(+)}$ both
contribute to the parameter b of the three-nucleon potential
and their resultant contribution is such that b is more or less
independent of $Z$. On the other hand, the parameter d varies appreciably
with $Z$. However, the contribution from d to the binding energy correction
$B_{3}$ of triton has been found to be much smaller compared with that
from b~\cite{Stadler,Ishikawa}. Therefore, the calculation of
$B_{3}$ is not likely to be quite sensitive to the variation of $Z$
as long as $Z$ is constrained to lie within its empirical
bounds, $|Z| \leq 1/2$.

The plan of the remaining portion of the paper is as follows: in Sec.\ 2
we review briefly the derivation of the $2\pi E-3NP$ using the
pion-nucleon off-shell scattering
amplitudes as input, in Sec.\ 3 we discuss our model for pion-nucleon
scattering and evaluate the nonrelativistic reduction of the amplitudes
which are to be used in the $2\pi E-3NP$. Finally, a detailed discussion
of the results are given in Sec.\ 4.


\section{Two-pion exchange three-nucleon potential }

A three-nucleon potential means an irreducible potential energy function
of the coordinates of the three nucleons --- irreducible in the sense that
the function cannot be written as a sum of functions involving fewer
coordinates. The Feynman diagram corresponding to $2\pi E-3NP$ is
shown in figure 1.
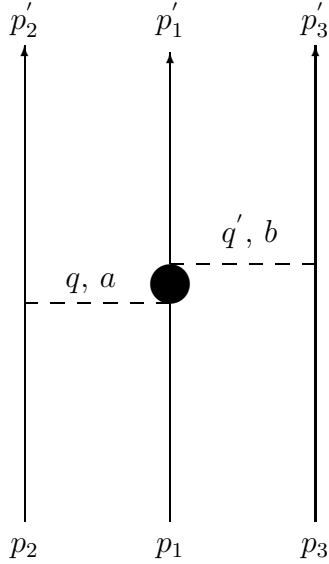
\begin{figure}
  \begin{center}
    \begin{picture}(100,200)
      \put(60,100){\circle*{15}}
      \put(60,10){\line(0,1){82.5}}
      \put(60,107.5){\vector(0,1){80}}
      \multiput(5,10)(110,0){2}{\vector(0,1){180}}
      \multiput(5,92.5)(10,0){6}{\line(1,0){5}}
      \multiput(60,107.5)(10,0){6}{\line(1,0){5}}
      \put(5,0){\makebox(0,0){$p_{2}$}}
      \put(60,0){\makebox(0,0){$p_{1}$}}
      \put(115,0){\makebox(0,0){$p_{3}$}}
      \put(5,200){\makebox(0,0){$p_{2}^{'}$}}
      \put(60,200){\makebox(0,0){$p_{1}^{'}$}}
      \put(115,200){\makebox(0,0){$p_{3}^{'}$}}
      \put(30,100){\makebox(0,0){$q,\,a$}}
      \put(90,120){\makebox(0,0){$q^{'},\,b$}}
    \end{picture}
  \end{center}
  \caption{ Two-pion exchange three-nucleon potential. The blob
           represents the pion-nucleon interaction}
\end{figure}
The amplitude for the process shown in figure 1 can be written as
\be
<p^{\prime}_{1}p^{\prime}_{2}p^{\prime}_{3} \mid S-1 \mid p_{1}p_{2}p_{3}>
\, = -i \delta^{4} \left ( P - P^{\prime} \right )\frac{1}{(2\pi)^{5}}
\sqrt{ \frac{m^{6}}{p_{10}p_{20}p_{30}p^{\prime}_{10}p^{\prime}_{20}
p^{\prime}_{30} } }\, \, T^{3N}_{123} ,
\ee
where
\be
T^{3N}_{123}  =  \left [ \bar{u}(p^{\prime}_{2})\gamma^{\mu}q_{\mu}
\gamma_{5}\tau_{a}u(p_{2}) \right ] \frac{f/\mu}{q^{2}-\mu^{2}}
\left \{ T^{ba}_{\pi N} \right \} 
\frac{f/\mu}{q^{\prime \, 2}-\mu^{2}} \left [
\bar{u}(p_{3}^{\prime})\gamma^{\nu}q^{\prime}_{\nu}
\gamma_{5}\tau_{b}u(p_{3}) \right ].
\ee
The pseudovector coupling for the $\pi NN$ vertex has been chosen in
conformity with the results of the nonlinear realization of the chiral
$SU(2) \times SU(2)$ symmetry for pion-nucleon interaction~\cite{Weinberg1}.

In the expressions (1) and (2) $P$ and $P^{\prime}$ are the total four
momenta before and after scattering; $q=p_{2}-p^{\prime}_{2}$ and
$ q^{\prime}= p^{\prime}_{3}-p_{3}$; a and b are the isospin indices of the
pion, and $\mu$ is the mass. The off-shell pion-nucleon T-matrix
$T_{\pi N}^{ba}$ describes the scattering process
\be
\pi^{a}(q) + N(p_{1}) = \pi^{b}(q^{\prime}) + N(p_{1}^{\prime}).
\ee
The pion-nucleon T-matrix is related to the S-matrix through the relation
\be
<q^{\prime}p^{\prime}_{1} \mid S_{\pi N}-1 \mid qp> = -i(2\pi)^{4}
\delta^{4} \left (q+p_{1}-q^{\prime}-p_{1}^{\prime} \right )
\sqrt{ \frac{m^{2}}{p_{10}p_{10}^{\prime}q_{0}q_{0}^{\prime}} } \,
T_{\pi N}^{ba} (\nu , t, q^{2},q^{\prime 2}),
\ee
where $\nu$ and $t$ are defined as
\be
 \nu = \frac{(q+q^{\prime}).(p_{1}+p_{1}^{\prime})}{4m},
\ee
\be
 t = (q-q^{\prime})^{2}.
\ee
The quantity $\nu$ is related to the Mandelstam variables
$s=(q+p_{1})^{2}$ and $u=(p_{1}-q^{\prime})^{2}$ by
$\nu = (s-u)/4m$. Now the T-matrix $T_{\pi N}^{ba}$ has the general
isospin decomposition
\barr
T_{\pi N}^{ba} & = & \bar{u}(p_{1}^{\prime}) \left \{ \left [ A^{(+)} +
\frac{1}{2} (\not\!{q} + \not\!{q}^{\, \prime})B^{(+)} \right ]
\delta_{ba} + \right. \nonumber \\
 &    & \left. \left [ A^{(-)}+\frac{1}{2} (\not\!{q}+\not\!{q}^{\, \prime})
  B^{(-)} \right ] i \epsilon_{bac}\tau_{c} \right \} u(p_{1}),
\earr
where $A^{(\pm)}$ and $B^{(\pm)}$ are the isospin-even(+) and the
isospin-odd(-) invariant amplitudes. An alternative isospin decomposition of
$T_{\pi N}^{ba}$ is
\be
 T_{\pi N}^{ba} = \bar{u}(p_{1}^{\prime}) \left \{ \left ( F^{(+)} -
 \frac{[\,\not\!{q}^{\, \prime}, \not\!{q}\,]}{4m}B^{(+)} \right ) \delta_{ba} +
 \left ( F^{(-)}
 - \frac{[\,\not\!{q}^{\, \prime}, \not\!{q}\,]}{4m}B^{(-)} \right ) i \epsilon_{bac} \tau_{c}
  \right \} u(p_{1}),
\ee
where
\be
 F^{(\pm)} = A^{(\pm)} + \nu B^{(\pm)}.
\ee
Now, since the potential is a nonrelativistic concept, we need to take the
nonrelativistic limit of Eq.\ (1) to define a three-nucleon potential.
In the theory of nonrelativistic potential scattering the S-matrix and the
T-matrix are related by
\be
<{\vec{p}_{1}}^{\;\, \prime } {\vec{p}_{2}}^{\;\, \prime} {\vec{p}_{3}}^{\;\,
 \prime} \mid s-1 \mid {\vec{p}_{1}} {\vec{p}_{2}} {\vec{p}_{3}}>\, =
-2\pi i \delta(E-E^{\prime})
<{\vec{p}_{1}}^{\;\, \prime} {\vec{p}_{2}}^{\;\, \prime} {\vec{p}_{3}}^
{\;\, \prime} \mid t
\mid {\vec{p}_{1}} {\vec{p}_{2}} {\vec{p}_{3}}>.
\ee
In the first approximation the t-matrix in Eq.\ (10) can be equated to the
three-nucleon potential W. The nonrelativistic reduction of Eq.\ (1) can
be compared to Eq.\ (10) to obtain the 3NP. Thus we find
\be
<{\vec{p}_{1}}^{\;\, \prime} {\vec{p}_{2}}^{\;\, \prime} {\vec{p}_{3}}^
{\;\, \prime} \mid W(123) \mid
{\vec{p}_{1}} {\vec{p}_{2}} {\vec{p}_{3}}> \, \, \approx \,\,
- \frac{1}{(2\pi)^{6}} \delta^{3}(\vec{P}-{\vec{P}}^{\, \prime})
t^{3N}_{123}\, \, ,
\ee
where $t^{3N}_{123}$ is the nonrelativistic reduction of $T^{3N}_{123}$.
Taking the appropriate limit, we finally obtain
\barr
\lefteqn{ <{\vec{p}_{1}}^{\;\, \prime} {\vec{p}_{2}}^{\;\, \prime} {\vec{p}_{3}}^
{\;\, \prime} \mid W(123) \mid {\vec{p}_{1}} {\vec{p}_{2}} {\vec{p}_{3}}>
\, =}  \nonumber \\
 & & \hspace{-0.3cm} - \, \frac{1} {(2\pi)^{6}} \delta^{3}(\vec{P}-{\vec{P}}^{\, \prime}) \left
( \frac{f}{\mu} \right )^{2} \frac{H({\vec{q}}^{\; 2})}{{\vec{q}}^{\; 2} +
\mu^{2}} \frac{H({\vec{q}}^{\; \prime \, 2})}{{\vec{q}}^{\; \prime \, 2} +
\mu^{2}} (\vec{ \sigma}_{2}. \vec{q})(\vec{\sigma}_{3}.{\vec{q}}^{\; \prime})
\tau_{a}^{(2)}\tau_{b}^{(3)} \times  \nonumber \\
 & & \hspace{-0.3cm} \left \{ \left [ f^{(+)} -  \frac{i}{2m}
\vec{\sigma}_{1}. (\vec{q} \times {\vec{q}}^{\; \prime})b^{(+)}
\right ] \delta_{ba}   
 +  \left [ f^{(-)} - \frac{i}{2m} \vec{\sigma}_{1}.
(\vec{q} \times {\vec{q}}^{\; \prime})b^{(-)} \right ] i
\epsilon_{bac} \tau_{c}^{(1)} \right \} \!\! ,
\earr
where $f^{(\pm)}$ and $b^{(\pm)}$ are the nonrelativistic limits of
$F^{(\pm)}$ and $B^{(\pm)}$; $H(\vec{q}^{\; 2})$ and
$H({\vec{q}}^{\; \prime \, 2})$
refer to the form factors which are introduced because the pions
are off-shell. We take $H({\vec{q}}^{\;2})$ as
\be
H(\vec{q}^{\,2}) = \left ( \frac{ \Lambda^{2} - \mu^{2}}{
\Lambda^{2} + \vec{q}^{\; 2}} \right )^{2}.
\ee

%
\section{Model for pion-nucleon interaction}
\subsection{The Weinberg Lagrangian}

We begin with the Weinberg Lagrangian \cite{Weinberg1} which is based on a
nonlinear realization of the chiral $SU(2) \times SU(2)$ symmetry.
The interaction Lagrangian relevant for pion-nucleon scattering can be
written as
\be
  {\cal L}_{W} = {\cal L}_{\pi NN} + {\cal L}_{\pi \pi NN}
\ee
where
\be
  \label{eq:deri}
  {\cal L}_{\pi NN} = (f/\mu) \bar{\psi}\gamma_{5} \gamma^{\mu}
  \tau_{i} \psi \partial_{\mu} \phi_{i},
\ee
\be
  \label{eq:dir}
  {\cal L}_{\pi \pi NN} = (i/4f_{\pi}^{2}) (\bar{\psi}i\gamma^{\mu}
  \tau_{i} \psi) \epsilon_{ijk}\phi_{j}\partial_{\mu}\phi_{k}.
\ee
Here $\psi$ and $\phi$ are the nucleon and the pion fields,
$f_{\pi} = 92.6$ MeV \cite{PDG} is the pion decay constant and $\mu$ the mass
of the pion. The interaction Lagrangian ${\cal L}_{W}$ consists of
the usual derivative pion-nucleon coupling (Eq.\ \ref{eq:deri}) and a direct
interaction between a pion and a nucleon (Eq.\ \ref{eq:dir}).

\subsection{The $\pi N \Delta$ interaction Lagrangian}

The most general form of the interaction Lagrangian
${\cal L}_{\pi N\Delta}$ can be written in the form~\cite{Nath}
\be
 \label{eq:Lagrangian}
 {\cal L}_{\pi N\Delta} = \frac{1}{\sqrt{2}} \left( \frac{f^{*}}{\mu} \right)
         \left [ i \overline{\Psi}_{\mu} \Theta^{\mu\nu} T_{i}\psi
         \partial_{\nu}\Phi_{i}  + h.c. \right ],
\ee
\be
  \label{eq:theta}
  \Theta_{\mu\nu} = \left \{ g_{\mu\nu} + \left[ \frac{1}{2} ( 1 + 4 Z ) A +
  Z \right] \gamma_{\mu} \gamma_{\nu} \right \},    
\ee
where $\Psi_{\mu}$ is the Rarita-Schwinger  field and the T's
are a set  of matrices  corresponding  to  the  isospin-$\frac{3}{2}.$
The propagator for the $\Delta (1232)$ is written as
\be
\langle 0|T( \psi_{\mu}(x)\bar{\psi}_{\nu}(y) )|0 \rangle \, =
        i d_{\mu \nu}(\partial)\Delta_{F}(x-y)
\ee
where
\barr
\label{eq:propagator}
\lefteqn{d_{\mu \nu}(\partial) =} \!\!\!\!\!\!\!  \nonumber \\
& & (i\gamma^{\lambda}\partial_{\lambda}+M) \left [ g_{\mu \nu}
-\frac{1}{3} \gamma_{\mu}\gamma_{\nu} - \frac{1}{3M}(\gamma_{\mu}
i\partial_{\nu} - \gamma_{\nu}i\partial_{\mu})
+\frac{2}{3M^{2}}\partial_{\mu}\partial_{\nu} \right ] +  \nonumber \\
& & \frac{1}{3M^{2}} \left ( \frac{A+1}{2A+1} \right )
\left \{ \left [- \frac{1}{2} \left ( \frac{A+1}{2A+1} \right )
i\gamma^{\lambda}\partial_{\lambda} + \left ( \frac{A}{2A+1} \right )M \right]
\gamma_{\mu}\gamma_{\nu} - \right. \nonumber \\
& & \gamma_{\mu}i\partial_{\nu} -
\left. \left ( \frac{A}{2A+1} \right ) \gamma_{\nu}i\partial_{\mu} \right \}
(\Box + M^{2})\,    
\earr
and
\be
  \Delta_{F}(x-y) = \frac{1}{(2\pi)^{4}}\int d^{4}p \,
  \frac{exp[-ip(x-y)]}{p^{2} - M^{2} +i\epsilon}\, .
\ee
The interaction Lagrangian ${\cal L}_{\pi N\Delta}$
depends on two parameters
A and $Z$. The parameter A, which occurs also in
the propagator, can assume any value except -1/2~\cite{Nath}.
However, A drops out from the final expressions of the scattering amplitudes
which therefore depend on $Z$ only. There is no consensus on the exact
value of $Z$, although $Z = 1/2$ is preferred theoretically~\cite{Nath}.
From phenomenological studies
a reliable bound,
\be
  \label{eq:bound}
  \mid Z \mid \leq \frac{1}{2},
\ee
can be placed on the value of $Z$ [5 - 9]. 
The value of the $\pi N\Delta$ coupling
constant is taken as $f^{*^{2}}/4\pi = 0.3359$. As the $\Delta(1232)$ makes the
dominant contribution to the $2\pi E-3NP$, and as there is some confusion
regarding the magnitude of the $\Delta(1232)$ contribution, we shall
discuss in detail this aspect of the problem in section 4.

\subsection{ The pion-nucleon $\sigma$-term}
\label{secsigma}
Current Algebra and PCAC impose certain constraints \makebox{[11 - 13]}
on the isospin-even invariant amplitude $A^{(+)}$ at some unphysical values of
the kinematical variables. To satisfy these constraints and to account for
the empirical information on the low-energy $\pi N$ scattering,
we include in our calculations an additional amplitude $A_{\sigma}^{(+)}$,
called the pion-nucleon $\sigma$-amplitude which is parametrized
\makebox{[6 - 8, 18]} as follows:
\be
\label{eq:sigma}
 A_{\sigma}^{(+)} ( \nu ,\nu_{B} ) = \frac{ \sigma_{NN} ( t = 2\mu^2) }
      { f_{\pi}^{2} } \left [ \frac{ q^{2} + q^{\prime 2} - \mu^{2} }{ \mu^{2}}
      + \frac{ \sigma^{\prime} (4m\nu_{B}) }{\mu^2}    \right ],
\ee
where
            
\[    \nu = (s-u)/4m, \,\,\,  \nu_{B} = (t-q^{2}-q^{\prime 2})/4m; \]
s, t and u are the Mandelstam variables, $q$ and $q^{\prime}$ are the momenta of the
incoming and the outgoing pion respectively. 

Recently the pion-nucleon scattering amplitudes in the subthreshold region
have been recalculated  by using the meson factory $\pi N$ data and
dispersion relation~\cite{sub}.
The important results relevant for our discussion are
\be
\label{eq:subthreshold}
\begin{array}{lcr}
 \bar{F}^{(+)}(\nu = 0, \, t = 2\mu^{2}, \,  q^{2} = \mu^{2}, \,
 q^{\prime 2} = \mu^{2}) & \approx &   1.35 \,\mu^{-1}    \\
 \bar{F}^{(+)}(\nu = 0,\, t = \mu^{2}, \,\, q^{2} = \mu^{2}, \,
 q^{\prime 2} = \mu^{2}) & \approx & -0.08 \,\mu^{-1}   \\   
 \bar{F}^{(+)}(\nu = 0, \, t = 0, \,\,q^{2} = \mu^{2}, \,
 q^{\prime 2} = \mu^{2}) & \approx &  -1.34 \,\mu^{-1}   
\end{array}
\ee
which yield
\be
 \sigma_{NN} (t=2\mu^{2}) = 82 \, {\rm MeV}
\ee
and
\be
\label{eq:sigp}
\sigma^{'} = \left \{  \begin{array}{ll}
                           0.66 & \mbox{for Z = 1/2}  \\
                           0.50 & \mbox{for Z = 1/4}  \\
                           0.40 & \mbox{for Z = 0}    \\
                           0.36 & \mbox{for Z = - 1/4}  \\
                           0.37 & \mbox{for Z = -1/2\, .}
                           \end{array}
             \right .
\ee
The amplitude $\bar{F}^{(+)}$ is the remainder of $F^{(+)}$ after the
(pseudovector) nucleon Born terms have been subtracted from it.
It may be noted here that while the value of $\sigma^{\prime}$ is sensitive
to the choice of the parameter $Z$ in the $\pi N\Delta$ interaction
Lagrangian, $\sigma_{NN}(t = 2\mu^{2})$ is independent of $Z$.

\subsection{Nonrelativistic limits of pion-nucleon scattering amplitudes}

The contributions to the pion-nucleon invariant amplitudes $A^{(\pm)}$
and $B^{(\pm)}$ due to nucleon-exchange, $\Delta$-exchange and direct
$\pi \pi NN$ interaction can be easily calculated and are quoted in
different places \cite{Nath,Shamsun}.
For our purpose we need only the
nonrelativistic reductions $f^{(\pm)}$ and $b^{(\pm)}$ of the amplitudes
$F^{(\pm)} = A^{(\pm)}+\nu B^{(\pm)}$ and $B^{(\pm)}$ respectively.

First, consider the nucleon-exchange contribution to pion-nucleon scattering.
The invariant amplitudes $A_{N}^{(\pm)}$ and $B_{N}^{(\pm)}$ consists of
the forward propagating Born term (FPBT) and the backward propagating Born
term (BPBT). The FPBT is already accounted for as the iterate of
two-nucleon one-pion exchange potential. Therefore the FPBT has to be
subtracted from the invariant amplitudes. If we take the nonrelativistic
limit of what remains we obtain 
\be
  \label{eq:n}
  f_{N}^{(\pm)} =0 \,\,\, ,\,\,\, b_{N}^{(\pm)} =0 \, .
\ee
We thus see that the nucleon-exchange contribution to the $\pi N$ amplitudes
does not contribute to the $2\pi E-3NP$. The results in Eq.\ (\ref{eq:n})
are correct only if we use the gradient coupling for $\pi NN$ interaction.

For the $\Delta$-contribution to the $\pi N$ amplitudes, we find
\barr
 F_{\Delta}^{(+)}\rightarrow & f_{\Delta}^{(+)} &= \alpha_{\Delta}^{(+)}
 \vec{q}.{\vec{q}}^{\; \prime}, \nonumber \\
 F_{\Delta}^{(-)}\rightarrow &f_{\Delta}^{(-)} &=0,
\earr
and
\barr
  B_{\Delta}^{(+)} \rightarrow & b_{\Delta}^{(+)} & =0, \nonumber \\
  B_{\Delta}^{(-)} \rightarrow & b_{\Delta}^{(-)} & = \beta_{\Delta}^{(-)}(Z),
\earr
where
\barr
  \label{eq:alpha}
  & \alpha_{\Delta}^{(+)} = &\left ( 2f^{*^2}/9\mu^{2} \right )
                    \left [ \frac{(4M^{2}-Mm+m^{2})}{(M-m)M^{2}}
                     \right. \nonumber  \\
         & & \left. \mbox{} -\frac{4(M+m)Z}{M^{2}}-\frac{4(2M+m)Z^{2}}{M^{2}}
         \right ]
\earr
and
\barr
  \label{eq:beta}
  & \beta_{\Delta}^{(-)} = & \left ( f^{*^{2}}/ 9\mu^{2} \right )
  \left [\frac{2m(2M^{2}+Mm-m^{2})}{(M-m)M^{2}} \right. \nonumber  \\
  & & \left. \mbox{} + \frac{8m(M+m)Z}{M^{2}} +
  \frac{8m(2M+m)Z^{2}}{M^{2}} \right ].
\earr
Here $ M=1232$ MeV is the mass of $\Delta(1232)$ and $ m= 938.9$ MeV is the
nucleon mass. The quantities $\alpha_{\Delta}^{(+)}$ and
$\beta_{\Delta}^{(-)}$ are obtained from the expressions for the
$\Delta$-contribution to the amplitudes $A^{(\pm)}$ and $B^{(\pm)}$ as given 
in Ref.\ \cite{Shamsun}.
These results (Eqs.\ \ref{eq:alpha}, \ref{eq:beta}) are the same as
derived earlier by Coelho, Das and Robilotta~\cite{Brazil}.
However, they chose $Z = -1/2$ for the detailed discussions on the
$\Delta$-contribution to the three-nucleon potential.

Next, for the direct $\pi \pi NN$ interaction, only the amplitude
$B_{d}^{(-)}=1/2f_{\pi}^{2}$ is nonzero. We therefore have
\be
  b_{d}^{(+)} = 0,\,\,\, b_{d}^{(-)} = 1/2f_{\pi}^{2},
\ee
and
\be
  f_{d}^{(+)}=0, \,\,\, f_{d}^{(-)} = \nu /2f_{\pi}^{2} \approx 0.
\ee
Finally, the nonrelativistic limit of the $\sigma$-contribution to the
$\pi N$ amplitude is also simple. We have
\be
  f_{\sigma}^{(+)} = a_{\sigma}^{(+)} = \frac{\sigma_{NN}(t = 2\mu^{2})}
  {f_{\pi}^{2}} \left [ -1 + \frac{ 2\sigma^{\prime}
  \vec{q}.\vec{q}^{\; \prime}}{\mu^{2}}
  -\frac{(\vec{q}^{\;2}+\vec{q}^{\; \prime \; 2})}{\mu^2} \right ].
\ee
Since the pion is off-shell, each of the pion-nucleon amplitudes have to be
multiplied by the form factor given in Eq.\ (13). Now, inserting the separate
contributions to $f^{(\pm)}$ and $b^{(\pm)}$ in Eq.\ (12) we can write the
$2\pi E-3NP$ as
\barr
\label{eq:3}
\lefteqn{<{\vec{p}_{1}}^{\;\, \prime } {\vec{p}_{2}}^{\;\, \prime} {\vec{p}_{3}}^{\;\,
 \prime} \mid W(123) \mid {\vec{p}_{1}} {\vec{p}_{2}} {\vec{p}_{3}}>\,  = }
 \nonumber \\
 & & \frac{1}{(2\pi)^{6}} \delta^{3}(\vec{P}-\vec{P}^{\, \prime})
 \left ( \frac{f}{\mu} \right )^{2}
 \frac{H({\vec{q}}^{\; 2})}{({\vec{q}}^{\; 2} +\mu^{2})}
 \frac{H({\vec{q}}^{\; \prime \; 2})}{({\vec{q}}^{\;\prime \; 2} +\mu^{2})}
 \,(\vec{\sigma}_{2}.\vec{q})(\vec{\sigma}_{3}.\vec{q}^{\;\prime}) \times
 \nonumber  \\
 & & \left \{ \vec{\tau}_{2}.\vec{\tau}_{3} \left [ a +
 b \vec{q}.{\vec{q}}^{\; \prime} + c({\vec{q}}^{\; 2}+
 {\vec{q}}^{\; \prime \;2}) \right ] - d(\vec{\tau}_{1}.\vec{\tau}_{2} \times \vec{\tau}_{3})
 (\vec{\sigma}_{1}. \vec{q} \times {\vec{q}}^{\; \prime}) \right \},
\earr
where
\barr
  a & = & \sigma_{NN}(t=2\mu^{2})/f_{\pi}^{2} \nonumber \\
  b & = & -\alpha_{\Delta}^{(+)}(Z) - 2\sigma^{\prime} \frac{
          \sigma_{NN}(t=2\mu^{2})}{f_{\pi}^{2}\mu^{2}} \nonumber \\
  c & = & \frac{\sigma_{NN}(t=2\mu^{2})}{f_{\pi}^{2} \mu^{2}} \nonumber \\
  d & = & - \frac{\beta_{\Delta}^{(-)}(Z)}{2m} - \frac{1}
          {4mf_{\pi}^{2}} \; .
\label{eq:3np}
\earr
The $2\pi E-3NP$ given in Eq.\ (\ref{eq:3}) is of the same form as
that derived by the
Tucson-Melbourne (TM) group 
\cite{TM}
except that the coefficients a,\,b,\,c and d in
the TM potential are:
\be
\begin{array}{rrr}
  a & = &  1.130  \, \mu^{-1}  \\
  b & = & -2.580  \, \mu^{-3}  \\
  c & = &  1.000  \, \mu^{-3}  \\
  d & = & \,-0.753  \, \mu^{-3}.  
  \end{array}
\label{eq:TM}
\ee

\section{Results and conclusions}

The parameters a, b, c and d in Eqs.\ (\ref{eq:3np}) receive contributions
from $A^{(+)}_{\sigma}$, the $\Delta$-exchange and the direct term for
$\pi N$ scattering. The $\Delta$-exchange contributes to b and d,
while the direct $\pi N$ interaction only to d. The parameters a and c
receive contributions from $A^{(+)}_{\sigma}$ only; $A^{(+)}_{\sigma}$
contributes also to b.

In order to show the relative importance of the various contributions, we
refer to table 1
where we have shown the values of a, b, c and d
corresponding to five different values of $Z$, namely $Z$ = 1/2, 1/4, 0,
-1/4 and -1/2. Note that a and c are independent of $Z$, but b and d are
not. Regarding the parameter b, as $Z$ is decreased from 1/2 to -1/4, the
contribution $b_{\Delta}$ from the $\Delta$-exchange decreases, while the
contribution $b_{\sigma}$ from $A^{(+)}_{\sigma}$ increases. However, this
trend is reversed somewhat at $Z$ = -1/2. As a result, the net b is not very
sensitive to the variation of $Z$ within its acceptable bounds,
$|Z| \leq 1/2$. Note that the $Z$-dependence of $b_{\sigma}$
is due to $\sigma^{\prime}$ which depends on $Z$ (Eq. \ref{eq:sigp}). Next,
the parameter d receives a small $Z$-independent contribution from the direct $\pi N$
scattering term, while the dominant contribution to d comes from the
$\Delta$-exchange which depends on $Z$. The value of d increases steadily
as $Z$ is decreased from 1/2 to -1/4, then it decreases
slightly at $Z$ = -1/2 (table 1).

Also shown in table 1 is the ratio $b_{\Delta}/d_{\Delta}$ which ranges from
1.26 to 4.0 as $Z$ is varied from 1/2 to -1/2. This contradicts the often
quoted result that $b_{\Delta}/d_{\Delta}$ should be equal to
four as a rule~\cite{chpert2}. In our calculations this ratio is
four only if $Z = -1/2$. However, there is no {\em a priori} justification
for choosing this value of $Z$. In fact, in the theory of spin-3/2 field
\cite{Nath,OO1}, $Z$ = -1/2 corresponds to calculations with
a $\pi N \Delta$ vertex and a\/ $\Delta$-propagator taking the
$\Delta$ on-mass-shell in both cases. More explicitly, if we take $A = -1$
in Eqs.\ (\ref{eq:theta}, \ref{eq:propagator}) and then $Z = -1/2$ in
Eq.\ (\ref{eq:theta}), the off-mass-shell parts of $\Delta$ are
eliminated from both the propagator and the interaction Lagrangian.
Peccei~\cite{Peccei} obtained a special form for the interaction
Lagrangian ${\cal L}_{\pi N\Delta}$ which would correspond to
$Z = -1/4$ in our formalism. For this value of $Z$, the ratio
$b_{\Delta}/d_{\Delta}$ is 4.23. However, if we take $Z = 1/2$,
the theoretically preferred value~\cite{Nath}, this ratio is 1.26,
much smaller than 4. It has already been noted in section (3.2) that,
while A may be assigned any value except $A = -1/2$, the
empirical bounds on $Z$ is $\mid Z \mid \leq 1/2$.

Our main purpose in this paper is to see in what way and to what extent the 
$Z$-dependence of the $\pi N\Delta $ interaction Lagrangian effects the
two-pion-exchange three-nucleon potential and whether any sensitive
$Z$-dependence is likely to appear in calculations of relevant physical
quantities, for example, the binding energy of triton.

A first-order perturbation calculation for the correction $E_3$ to the energy
of triton due to the Tucson-Melbourne potential with the parameters a, b, c
and d as in Eqs. (\ref{eq:TM}) was done by
Ishikawa {\it et al\/}~\cite{Ishikawa}. The
zeroth-order triton wave function was obtained by solving the Faddeev
equations with a variety of two-nucleon potentials. Ishikawa {\it et al\/}
used the dipole form factor at the vertices with several values of the
cut-off parameter $\Lambda $. For $\Lambda$ = 800 MeV and the Reid soft-core
two-nucleon potential they found that the contributions to $E_3$ from the
individual terms of the TM potential corresponding to the parameters a, b,
c, and d (Eqs. \ref{eq:TM}) are 0.05 MeV, -0.97 MeV, 0.25 MeV and -0.22 MeV
respectively. In the first-order calculations of Ishikawa {\it et al\/},
$E_{3}$ is linear in a, b, c and d. Since the two-pion exchange three-nucleon
potential in our model is of the same form as the TM potential, we can
easily estimate the binding energy correction $B_3(=-E_3)$ for triton due to
our $\pi \pi $-exchange three-nucleon potential, simply by scaling Ishikawa 
{\it et al\/}'s results (table 2). This is done solely to examine the
possible $Z$-dependence of $B_3$, although a first-order perturbative
calculation is not expected to be very accurate. We see that as $Z$
decreases from 1/2 to -1/2, $B_3^\Delta$ increases initially and then
remains more or less constant for negative $Z$, while $B_3^\sigma $
decreases and then becomes negligible for $Z\leq 0$. Therefore,
$B_3=B_3^\Delta +B_3^\sigma +B_3^d$, is not very sensitive to the
variation of $Z$. This conclusion would not change appreciably if the direct
$\pi N$ interaction is replaced by the $\rho$-mediated interaction.

\begin{table}[p]
\caption{The parameters a, b, c and d in the two-pion exchange
three-nucleon potential displayed for five values of $Z$ within the bounds
$|Z| \leq 1/2$. Note that a and c are independent of $Z$.}
\vspace{0.25cm}
\begin{center}
\begin{tabular}{|cc|c|ccc|ccc|c|} \hline \hline
$a = a_{\sigma}$ & $c = c_{\sigma}$ & Z & $b_{\Delta}$ & $b_{\sigma}$
& $b$ & $d_{\Delta}$ & $d_{d}$ &
$d$ & $b_{\Delta}/d_{\Delta}$ \\
($\mu^{-1}$) & ($\mu^{-3}$) & & ($\mu^{-3}$) & ($\mu^{-3}$) & ($\mu^{-3}$)
& ($\mu^{-3}$) & ($\mu^{-3}$) & ($\mu^{-3}$) & \\ \hline
1.341 & 1.341 &1/2 & -1.038 & -1.770 & -2.808 & -0.821 & -0.056 & -0.877 & 1.26 \\
 & & 1/4  & -1.445 & -1.341 & -2.786 & -0.618 & & -0.674 & 2.34 \\
 & & 0    & -1.706 & -1.073 & -2.779 & -0.487 & & -0.543 & 3.50 \\
 & & -1/4 & -1.820 & -0.966 & -2.786 & -0.430 & & -0.486 & 4.23 \\
 & & -1/2 & -1.787 & -0.992 & -2.779 & -0.447 & & -0.503 & 4.0  \\ \hline
\end{tabular}
\end{center}
\end{table}
\begin{table}[p]
\caption{The correction, $B_{3} = B_{3}^{\Delta} + B_{3}^{\sigma}
+ B_{3}^{d}$, to the binding energy of triton due to the
two-pion exchange three-nucleon potential. The contributions
$B_{3}^{\Delta}$ and $B_{3}^{\sigma}$ are $Z$-dependent. The table also
shows the binding energy $B_{2}$ of triton for the two-nucleon
Reid soft-core potential, the total
binding energy $B = B_{2}+B_{3}$ and $B_{exp}-B$. All the
binding energies are in MeV.}
\begin{center}
\begin{tabular}{|cc|ccccccc|} \hline \hline
$B_{2}$ & $B_{3}^{d}$ & Z & $B_{3}^{\Delta}$ & $B_{3}^{\sigma}$ &
$B_{3}$ & B & $B_{exp}$ & $B_{exp}-B$ \\ \hline
7.24 & 0.02 & 1/2  & 0.63 & 0.27  & 0.92 & 8.16 & 8.48 & 0.32  \\
          & & 1/4  & 0.72 & 0.11  & 0.85 & 8.09 & & 0.39        \\
          & & 0    & 0.78 & 0.01  & 0.81 & 8.05 & & 0.43         \\
          & & -1/4 & 0.81 & -0.03 & 0.80 & 8.04 & & 0.44          \\
          & & -1/2 & 0.80 & -0.02 & 0.80 & 8.04 & & 0.44   \\ \hline
\end{tabular}
\end{center}
\end{table}

It may be noted here that the numerical values for $B_{3}$ will depend on
the choice of the two-nucleon potential and the value of the cut-off
parameter $\Lambda$. In particular, the binding energy correction
is quite sensitive to $\Lambda$~\cite{Stadler,Ishikawa}. The dependence
on $\Lambda$ may be reduced somewhat if one includes the
$\rho \pi$-exchange, $\rho \rho$-exchange three-nucleon forces in
addition to the $\pi \pi$-exchange three-nucleon force~\cite{Stadler}.
However, our purpose is to investigate in detail the effect of
$\Delta(1232)$ on the parameters of the two-pion
exchange three-nucleon potential. The resonance $\Delta(1232)$ makes
large contributions to the parameters b and d, and these contributions
depend on $Z$. We find that $b_{\Delta}$ is sensitive to the variation
of $Z$, although $b = b_{\Delta}+b_{\sigma}$ does not change appreciably
with $Z$. The parameter d, however, depends substantially on $Z$.

\vspace*{2.0cm}

{\Large \bf Acknowledgement:}
We are grateful to the UGC, Bangladesh, for a financial grant supporting
this work.

\end{document}